\documentclass[letter]{aa}  

\usepackage{graphicx}
\usepackage{txfonts}
\usepackage{comment}
\usepackage{hyperref}
\newcommand{\eq}[1]{\begin{equation}  #1 \end{equation}}

\newcommand{\eqa}[1]{\begin{align}   #1 \end{align}}

\newcommand{\br}[1]{\left( #1 \right)}
\newcommand{\bc}[1]{\left\{ #1 \right\}}
\newcommand{\bb}[1]{\left[ #1 \right]}
\newcommand{\ba}[1]{\left\langle #1 \right\rangle}

\newcommand{\nn}{\nonumber}

\newcommand{\dd}{{\rm d}}

\newcommand{\expo}[1]{~{\rm e}^{ #1 }}

\newcommand{\pr}{{\rm Pr}}

\usepackage{cancel}
\DeclareMathOperator{\Tr}{Tr}

\begin{document}

 \title{When tension is just a fluctuation}
 \subtitle{How noisy data affect model comparison}
\titlerunning{When tension is just a fluctuation}

\author{B. Joachimi\inst{1},
F. K\"ohlinger\inst{2},
W. Handley\inst{3,4}
\and
P. Lemos\inst{1}
}

\institute{Department of Physics and Astronomy, University College London, Gower Street, London WC1E 6BT, UK\\
\email{b.joachimi@ucl.ac.uk}
\and
Kavli IPMU (WPI), UTIAS, The University of Tokyo, Kashiwa, Chiba 277-8583, Japan
\and
Astrophysics Group, Cavendish Laboratory, J.J. Thomson Avenue, Cambridge CB3 0HE, UK
\and
Kavli Institute for Cosmology, Madingley Road, Cambridge CB3 0HA, UK
}

\date{Received 2020 September 30; accepted 2021 February 10}

\abstract{Summary statistics of likelihood, such as Bayesian evidence,
offer a principled way of comparing models and assessing tension
between, or within, the results of physical experiments. Noisy realisations of the data induce scatter in these model comparison statistics. For a realistic case of cosmological inference
from large-scale structure, we show
that the logarithm of the Bayes factor attains scatter of order unity,
increasing significantly with stronger tension between the models under comparison. We develop an approximate
procedure that quantifies the sampling distribution of the evidence at
a small additional computational cost and apply it to real data to demonstrate the impact of the
scatter, which acts to reduce the significance of any model discrepancies. Data compression
is highlighted as a potential avenue to suppressing noise in the evidence to negligible levels, with a proof of concept demonstrated using \textit{Planck} cosmic microwave background data.}  

\keywords{Methods: data analysis -- Methods: statistical -- Cosmology: observations -- cosmic background radiation -- gravitational lensing: weak}

\maketitle

%

\section{Introduction}
\label{sec:intro}

\begin{figure*}
        \centering
        \includegraphics[width=2\columnwidth]{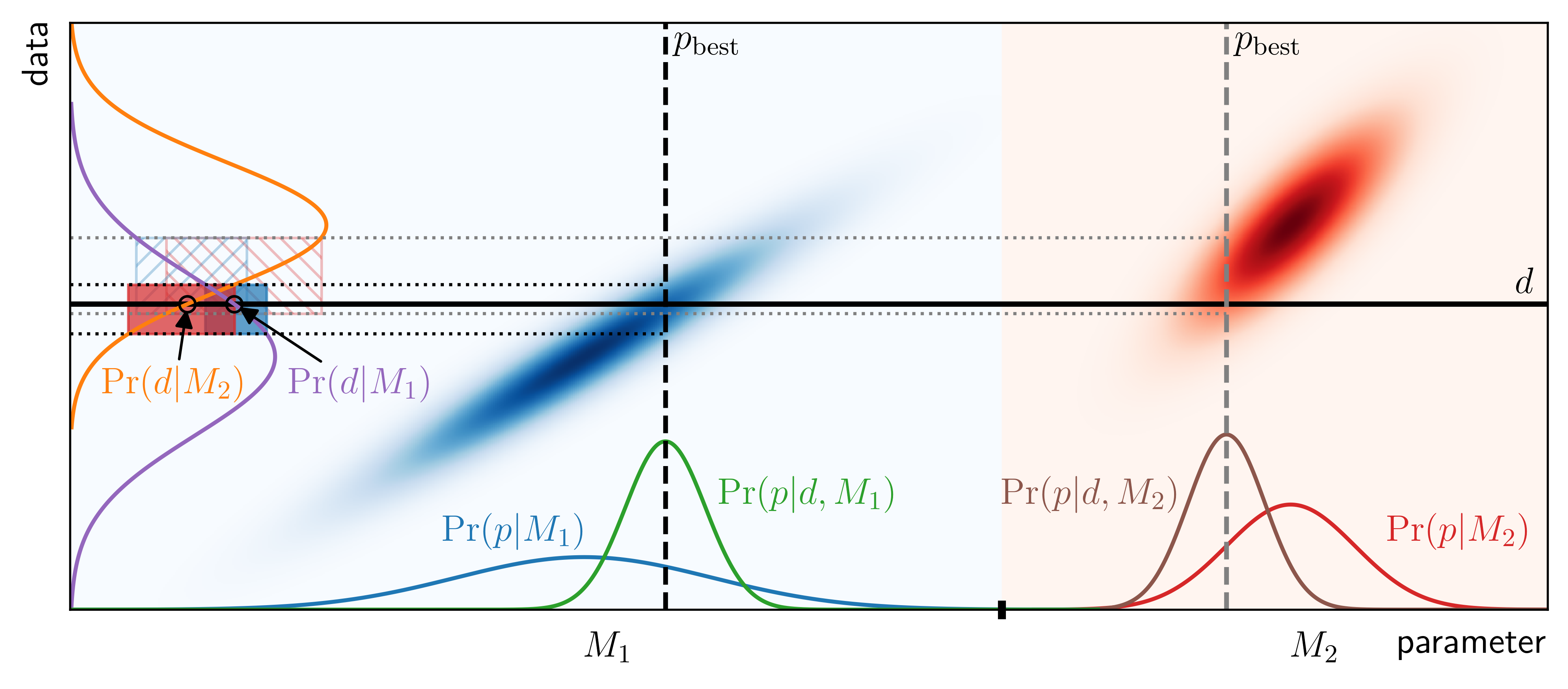}
    \caption{Illustration of model comparison via evidence and of its associated scatter for a one-dimensional data vector $d$ and two models, $M_i$ with $i=1,2$, each with a single parameter $p$. The joint distributions $\pr \br{d,p|M_i}$ are shown in blue and red shades. The projections of these distributions onto the parameter and data axes yield the prior $\pr \br{p|M_i}$ and the evidence or marginal likelihood, respectively (shown in purple and orange for $M_1$ and $M_2$). An experiment produces the observation $d$, shown as the solid black line. Conditioning on $d$ yields the posterior $\pr \br{p|d,M_i}$,  shown in green and brown for $M_1$ and $M_2$. Evaluating the marginal likelihood at $d$ yields $\pr \br{d|M_i}$, which is used in the model comparison. The dotted horizontal lines mark the $1\sigma$ interval of possible alternative realisations of the data given the best-fit parameter $p_{\rm best}$ (dashed vertical lines) of either model. The blue and red boxes show the resulting $1\sigma$ range in possible evidence values under $M_1$, which has higher evidence in this case (the corresponding ranges for $M_2$ are shown as hatched areas).}
        \label{fig:sketch}
\end{figure*}

Binary decisions inevitably have to be made at the conclusion of a physical experiment. These include whether or not a feature has been detected significantly, which model describes the data better, and whether datasets  (or subsets thereof) are consistent with each other or are in \lq tension\rq, which could be a potential indicator for new physics not incorporated in the model.

Traditionally, hypothesis tests were the statistical tools of choice to answer these questions. With the advent of high-performance computing, Bayesian techniques building on Bayesian evidence have risen in popularity (e.g. \citealp{jaffe96,kunz06,marshal06}; see \citealp{trotta08} for a review). In cosmology, discrepancies in the $\sim 3-5\sigma$ range between high-redshift observations -- primarily the cosmic microwave background (CMB), as constrained most accurately by the \textit{Planck} mission; \citealp{Planck2018} -- as well as various probes of the low-redshift universe in the measurement of the Hubble constant \citep[e.g.][]{riess16,riess18,riess19,wong19} and the amplitude of matter density fluctuations  \citep[e.g.][]{joudaki17,joudaki19,DES-Y1,asgari20,heymans20} have recently emerged  (cf. \citealp{verde19} for an overview). This has spurred a flurry of work on approaches to quantifying tension and performing model comparison \citep[e.g.][]{verde13,seehars14,lin17,charnock17,Koehlinger2019,handley19,nicola19,adhikari19,raveri19}.

What these methods have in common is that they infer a single scalar that is then compared against a predefined scale to judge significance. In Bayesian statistics, the tension measure is conditioned on the observed data. The posterior probability of the parameters, $\boldsymbol{p}$, of a model, $M_i$, for measured data, $\boldsymbol{d}$, is
\eq{
\pr (\boldsymbol{p} | \boldsymbol{d},M_i) = \frac{1}{{\cal Z}_i} \pr (\boldsymbol{d} | \boldsymbol{p},M_i)\, \pr(\boldsymbol{p}|M_i)\;,
}
where $\pr (\boldsymbol{d} | \boldsymbol{p},M_i)$ is the likelihood and $\pr(\boldsymbol{p}|M_i)$ the prior on the model parameters. The Bayesian evidence, or marginal likelihood, is the normalisation given by
\eq{
\label{eq:defevidence}
{\cal Z}_i \equiv \pr(\boldsymbol{d} | M_i) = \int \dd^m p\; \pr (\boldsymbol{d} | \boldsymbol{p},M_i)\, \pr(\boldsymbol{p}|M_i)\;,
}
where $m$ denotes the number of parameters. For a given dataset, ${\cal Z}_i$ reduces to a non-stochastic scalar that attains larger values the more likely the realisation of the data is under model $M_i$, and the more predictive the model is (as a more flexible model could accommodate many possible forms of the data).

However, a physical experiment generally does not take acquired data as a given, but rather interprets them as a stochastic realisation of an underlying truth that we wish to approximate by our model. A different realisation of the data leads to a different value for ${\cal Z}_i$, which could alter our decision on tension or consistency. In this view, statistical uncertainty in the data turns the evidence (or any related tension and model comparison measure) into a noisy statistic\footnote{This viewpoint will require us to go beyond a purely Bayesian approach. However, hybrid Bayesian-frequentist methods are commonplace in statistics; see \citet{good92} for an overview, as well as \citet{jenkins14}.}. \citet{jenkins11} argued, based on toy experiments and analytical arguments, that the thus-inherited statistical uncertainty in ${\cal Z}_i$ is substantial. Ignoring this scatter will therefore lead to over-confident or incorrect decisions in model comparison.

In this work we quantify the scatter in the Bayesian evidence and some of its derived tension or model comparison statistics, affirming the findings of \citet{jenkins11} in a realistic cosmological experiment. We devise a computationally efficient procedure to calculate statistical errors on the evidence, apply it to an analysis of internal consistency in Kilo Degree Survey (KiDS) weak lensing data, and explore the impact of data compression on evidence scatter using \textit{Planck} CMB data as an example.


\section{Noisy model comparison}
\label{sec:inf_model_comp}

Figure~\ref{fig:sketch} illustrates the notion of evidence and its associated scatter using a Gaussian toy model that is one-dimensional in both data and parameter space. It builds on Fig.~28.6 of \citet{mackay03}. While at the observed data Model 1 has higher evidence in this example, it is not unambiguously superior because alternative realisations of the data under the more probable Model 1 could result in equal or reversed evidence values of Models 1 and 2 instead (see the boxes in blue and red shading)\footnote{For ease of illustration, the toy model considers the likelihood of the data conditioned on the best-fit parameter $p_{\rm best}$; in our implementation we take the full posterior into account when drawing new data realisations.}. We seek to quantify this statistical uncertainty of the evidence (see also Appendix~\ref{app:gaussian_analytic} for a closed-form analytic calculation in the Gaussian case analogous to Fig.~\ref{fig:sketch}).

\subsection{Scatter in the evidence and the Bayes factor}

The standard statistic to compare two models, $i$ and $j$, is the Bayes factor (see \citealp{kass95} for a review),
\eq{
\label{eq:defbayesfactor}
R_{ij} \equiv \frac{\pr(M_i|\boldsymbol{d})}{\pr(M_j|\boldsymbol{d})} = \frac{{\cal Z}_i\, \pr(M_i)}{{\cal Z}_j\, \pr(M_j)} = \frac{{\cal Z}_i}{{\cal Z}_j}\;,
}
which, for equal prior probabilities of the models themselves, is given by the ratio of the model evidence values. Here, $R_{ij}$ has the intuitive interpretation of betting odds in favour of model $i$ over $j$. We shall assume initially that we know the true underlying model, $M_{\rm true}$, including its parameters, $\boldsymbol{p}_{\rm true}$, that generates the data we observe, which need not coincide with those from either $M_i$ or $M_j$. Then the probability density of the Bayes factor is given by
\eq{
\label{eq:bayes_distribution}
\pr(R_{ij}|M_{\rm true}) = \int \dd^n d'\; \pr(R_{ij}|\boldsymbol{d}')\, \pr(\boldsymbol{d}'| M_{\rm true})\;, 
}
where $n$ is the dimension of the data vector, $\pr(\boldsymbol{d}| M_{\rm true})$ is the true likelihood of the data, and $\pr(R_{ij}|\boldsymbol{d}) $ is the distribution of $R_{ij}$ for a given dataset, which we shall assume to be deterministic. Hence, if the true likelihood is known, we can proceed as follows to create a distribution of $R_{ij}$: (i) generate samples of the data from the true likelihood, and (ii) calculate the Bayes factor for each sample according to Eqs. (\ref{eq:defevidence}) and (\ref{eq:defbayesfactor}).

As a realistic example, we chose a recent cosmological analysis of tomographic weak lensing measurements by the KiDS survey (KiDS-450; \citealp{Kuijken2015,Hildebrandt2017}). We worked with a simulated data vector that, like the real data, has size $n=130$ and depends in a highly non-linear way on seven model parameters (five cosmological parameters of a flat $\Lambda$CDM model, plus two parameters describing the astrophysical effects on the observables). It was assumed that the data have a Gaussian likelihood with a known and fixed covariance. To perform an internal consistency test, we created two copies of the parameter set and assigned one copy to the elements of the data vector that is dependent on tomographic bin no. 3 and the other copy to the remaining elements. The model comparison is then between the analysis with the original set of model parameters (Model 0) and that with the doubled parameter set (Model 1). Details regarding the methodology and analysis are available in \citet{Koehlinger2019} and Appendix$\,$\ref{app:kids} of this letter.

We generated 100 realisations of the data vector from the true likelihood, evaluated at a fiducial choice of the parameters. For each simulated data vector, we repeated a full nested sampling analysis of both models (0 and 1) and inferred the evidence (see Appendix \ref{app:kids} for an assessment of the robustness of the sampling algorithms). By default, we did not introduce any systematic shift into our simulated data; as such, strong concordance is expected as the outcome of the tension analysis.
\begin{figure}
        \centering
        \includegraphics[width=\columnwidth]{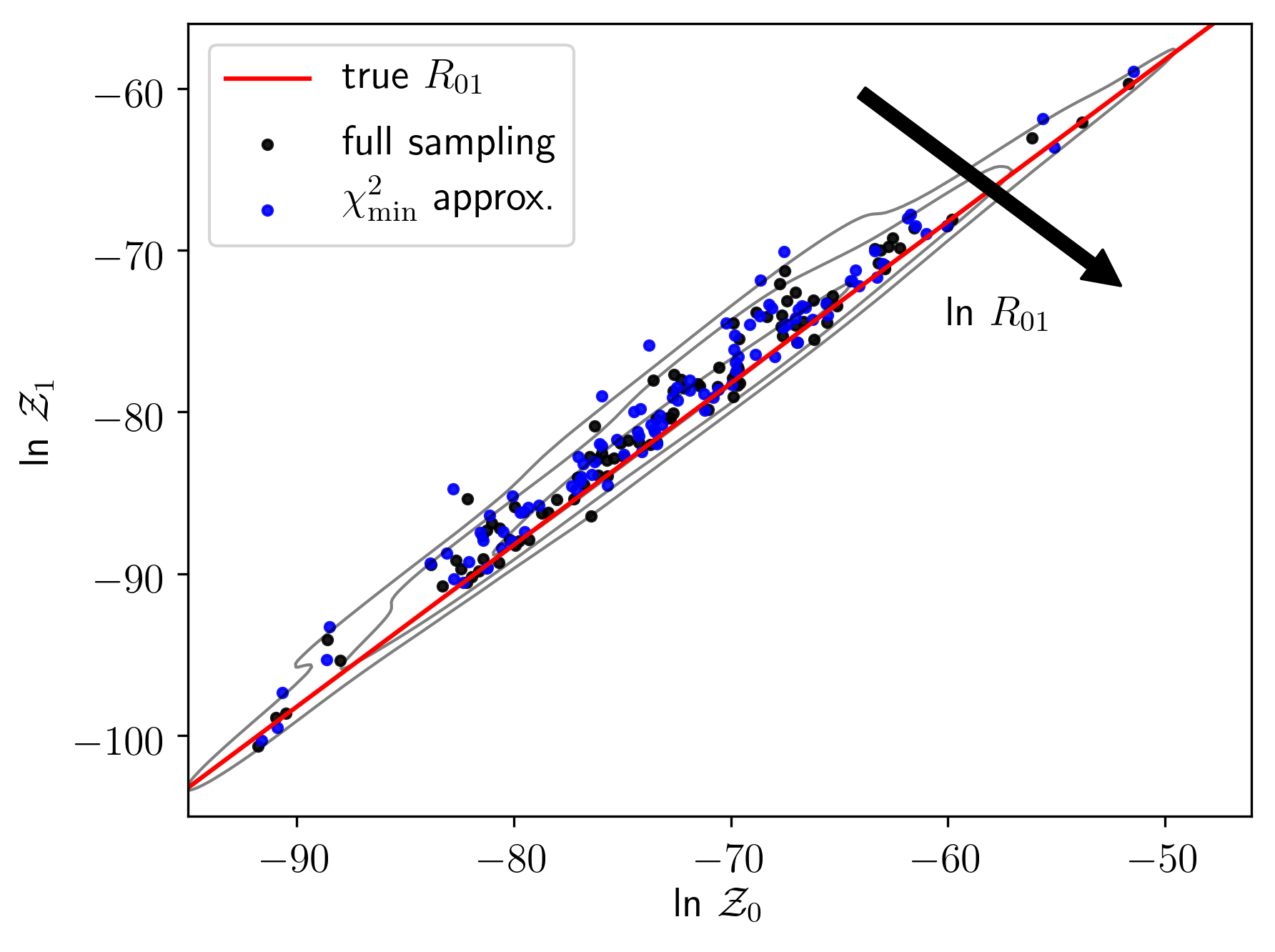}
    \caption{Joint distribution of the evidence calculated under two models (the mock KiDS cosmology analysis for a joint [Model 0] and a split [Model 1] data vector). The arrow indicates the direction of the increasing Bayes factor, $\ln R_{01}$, with the red line marking the true value of $R_{01}$. Black points correspond to the inferences from 100 noise realisations of the mock data vector, evaluated on the full nested sampling analysis, and the blue points correspond to the $\chi^2_{\rm min}$ approximation from Sect. \ref{sec:fastapprox}. Contours show the Gaussian kernel density approximation of the distribution based on the black points.}
        \label{fig:zsamples}
\end{figure}

The resulting distribution of evidence values is shown in Fig.$\,$\ref{fig:zsamples}. We computed the true value of the Bayes factor by re-running the analyses for a noise-free data vector generated for the fiducial parameter values. The two evidence distributions are each consistent with being lognormal\footnote{For a Gaussian likelihood, $\ln {\cal Z}$ is expected to be $\chi^2$-distributed.}, each with a standard deviation in the log of 7.9. The evidence is strongly correlated (Pearson correlation coefficient 0.99), which is plausible as the scatter derives from the same noisy data realisation, with both models yielding good fits.

\begin{figure}
        \centering
        \includegraphics[width=\columnwidth]{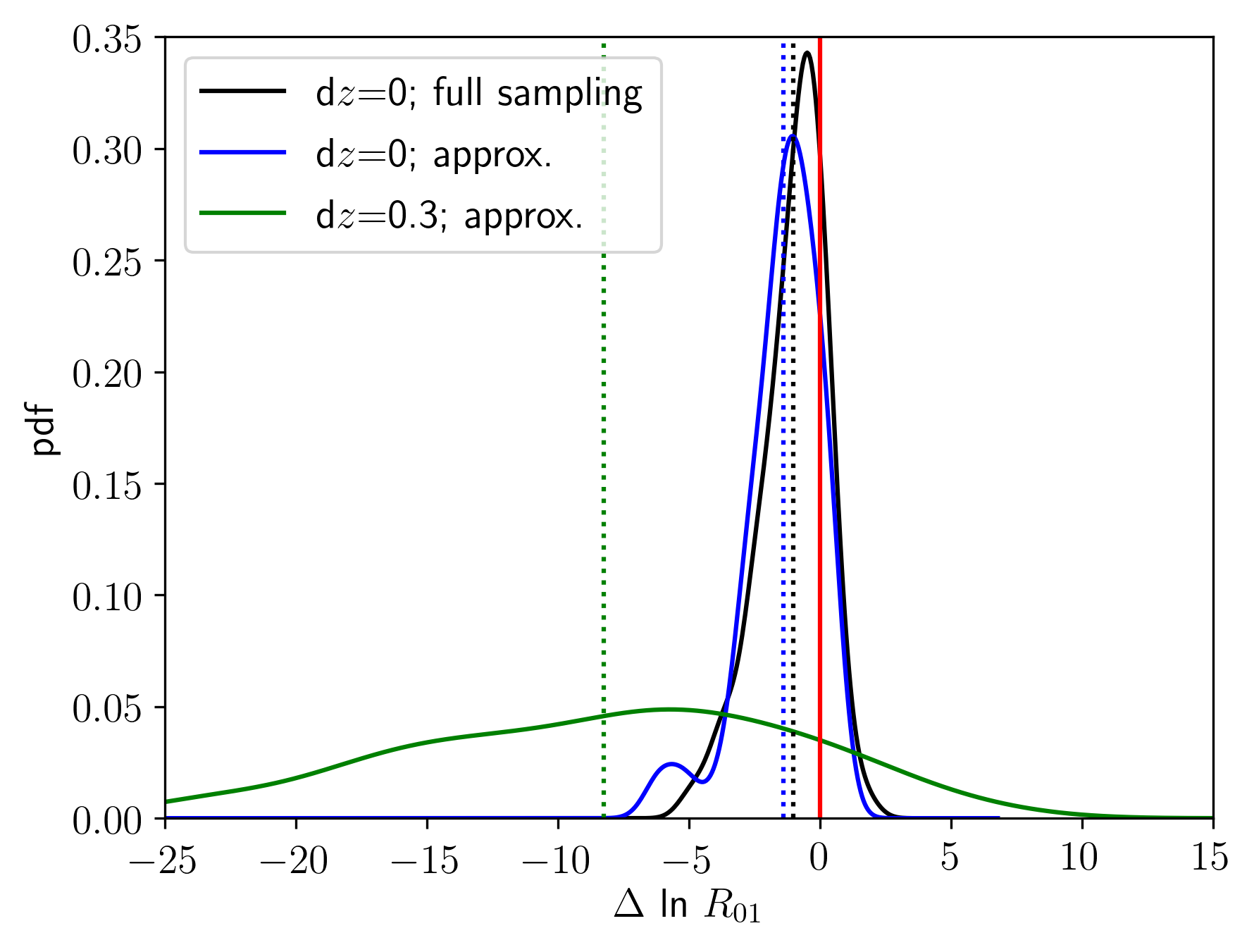}
    \caption{Probability density of the Bayes factor, $\ln R_{01}$, shifted to have its true value at 0. The black curve is the distribution of $\Delta \ln R_{01}$ extracted from the full nested sampling analysis, and the blue curve is the distribution extracted from the $\chi^2_{\rm min}$ approximation from Sect. \ref{sec:fastapprox}. The green curve corresponds to the highly discrepant case of one of four tomographic redshift bins being shifted by $\dd z=0.3$. Dotted lines mark the mean of each distribution.}
        \label{fig:rdistribution}
\end{figure}

Due to the strong correlation, the distribution of the Bayes factor is narrower, with $\sigma(\ln R_{01}) \approx 1.25$ (see Fig.$\,$\ref{fig:rdistribution} for its distribution). We also observe skewness in $\ln R_{01}$ (already visible in Fig.$\,$\ref{fig:zsamples}), which causes the mean to be lower relative to the true value by $\sim 1\sigma$. We do not find evidence that the skewness is due to numerical issues and so ascribe it to the non-linearity of the models; this means that this feature will be strongly dependent on the details of the analysis. A value of $\sigma(\ln R_{01}) \sim 1$ is in excellent agreement with the conclusions of \citet{jenkins11}, although they predicted a normal distribution for $\ln R_{01}$ (see also Appendix~\ref{app:gaussian_analytic}).

\subsection{Impact on suspiciousness}

By design, the Bayes factor depends on the parameter prior, which can be a hindrance for tension assessment, as demonstrated by \citet{handley19}. They proposed a modified statistic called \lq suspiciousness', defined as $\ln S_{ij} \equiv \ln R_{ij} + D_{{\rm KL},i} - D_{{\rm KL},j}$, where
\eq{
D_{{\rm KL},i} = \int \dd^m p\; \pr(\boldsymbol{p}|\boldsymbol{d},M_i)\; \ln \frac{\pr(\boldsymbol{p}|\boldsymbol{d},M_i)}{\pr(\boldsymbol{p},M_i)}\;
}
is the Kullback-Leibler (KL) divergence (\citealp{kl51}; see \citealp{lemos19} for a generalisation to correlated datasets, which we consider here). This combination of evidence and KL divergence is independent of the prior widths, providing they do not impinge upon the posterior bulk, and may be calibrated into a traditional \lq $\sigma$ tension\rq\ value.

Again assuming a Gaussian likelihood, $\ln S_{ij}$ is $\chi^2$-distributed with the degrees of freedom given by the difference in the effective dimension of the parameter space in the two models. \citet{handley19b} propose calculating this effective dimension as
\eq{
m_{{\rm eff},i} = 2 \bc{\ba{ \bb{ \ln \pr(\boldsymbol{p}|\boldsymbol{d},M_i) }^2}_{\rm p} -  \Big \langle \ln \pr(\boldsymbol{p}|\boldsymbol{d},M_i) \Big \rangle_{\rm p}^2}\;,
}
that is, twice the variance of the log-likelihood evaluated over the posterior distribution (indicated by the subscript \lq p').

We extracted $\ln S_{ij}$ from the output of our nested sampling analysis and determined the scatter of $m_{\rm eff}$ from the sub-sample variance computed on a posterior sample. The standard deviations of $\ln R_{01}$ and $\ln S_{01}$ agree to better than $10\,\%$; the following section outlines an argument for why the distributions of $R$ and $S$ are expected to be very similar. The standard deviation of $m_{\rm eff}$ is of the order of $10\,\%$ and can be treated as being uncorrelated with $S$ (correlation coefficient -0.17).

There are two obstacles to using the above approach on real data: (i) repeating full likelihood analyses that include evidence calculations many times to build a sample is prohibitively expensive, and (ii) we do not know the true likelihood to generate samples of the real data. We will address both these points in Sects.~\ref{sec:fastapprox} and \ref{sec:evidence_real}.

\subsection{Strong tension case}

To investigate a case of strong tension, we inserted a large shift, $\dd z = 0.3$, into the mean redshift of tomographic bin no. 3 of  the simulated data vectors and repeated the analysis\footnote{We employ the fast approximate algorithm detailed in Sect.$\,$\ref{sec:fastapprox}.}. In this case the alternative model 1 is clearly preferred ($\ln R_{01} \approx -23$). The impact on the distribution of $\ln R_{01}$ is dramatic, as can be seen from Fig.$\,$\ref{fig:rdistribution}. While the skewness and corresponding discrepancy between the mean and true values persist, the standard deviation increases to $7.3$, spanning more than three orders of magnitude in odds.

This result is driven by an increase in the scatter of the evidence (to 10.8 and 8.5 for models 0 and 1, respectively) and in particular by their partial de-correlation (correlation coefficient 0.74). As shown by \citet{jenkins11}, the scatter in $\ln R_{ij}$ is, to a good approximation, proportional to the typical difference between the model predictions at the respective best-fit parameters of the models under comparison (as also shown in Appendix~\ref{app:gaussian_analytic}). This difference is small in our concordant case with nested models, deviating from zero only through scatter in the data. In the $\dd z = 0.3$ case, the best fits of the models now lie far apart; this enlarges the scatter and propagates the noise differently into the evidence for models 0 and 1, thereby reducing their correlation.

\section{A fast approximate algorithm}
\label{sec:fastapprox}

We now consider the Laplace approximation of the log-likelihood (we dropped the explicit dependence on the model for simplicity),\footnote{We thank our referee for pointing out that this assumption has a more principled grounding, in that it maximises the entropy in the absence of further information on the form of the likelihood.} 
\eq{
\ln \pr(\boldsymbol{d}|\boldsymbol{p}) \approx \ln \pr(\boldsymbol{d}|\boldsymbol{p}_0) - \frac{1}{2} (\boldsymbol{p}-\boldsymbol{p}_0)^\tau\, \tens{F}(\boldsymbol{p}_0)\, (\boldsymbol{p}-\boldsymbol{p}_0) \;, 
}
where we expanded around the maximum of the log-likelihood at $\boldsymbol{p}_0$ and introduced the Fisher matrix,
\eq{
\tens{F}_{\alpha \beta} = - \ba{ \left. \frac{\partial^2 \ln \pr(\boldsymbol{d}|\boldsymbol{p})}{\partial p_\alpha\, \partial p_\beta} \right|_{\boldsymbol{p}_0} }\;,
}
as the negative expectation of the Hessian of the log-likelihood at $\boldsymbol{p}_0$. With this approximation the evidence reads
\eq{
{\cal Z} \approx \frac{(2 \pi)^{m/2}\, \pr(\boldsymbol{d}|\boldsymbol{p}_0)}{\sqrt{\det \tens{F}(\boldsymbol{p}_0)}\, V_{\rm prior}}\;,
}
where we additionally assumed that the prior is uninformative (i.e. the bulk of the likelihood lies well within the volume covered by the prior, denoted as $V_{\rm prior}$). Considering a Gaussian likelihood, such that $\pr(\boldsymbol{d}|\boldsymbol{p}_0)  \propto \expo{- \frac{1}{2} \chi^2(\boldsymbol{p}_0)}$, one finds \citep[cf.][]{handley19}
\eqa{
\label{eq:scatter_evidence}
\ln {\cal Z} &\approx \mbox{const.} - \ln V_{\rm prior} - \frac{1}{2} \ln \det \tens{F}(\boldsymbol{p}_0) - \frac{1}{2}\, \chi^2(\boldsymbol{p}_0)\; ~~~\mbox{and} \\ 
D_{\rm KL} &\approx \ln V_{\rm prior} - \frac{m}{2} \br{1 + \ln 2 \pi} + \frac{1}{2} \ln \det \tens{F}(\boldsymbol{p}_0)\;.
}
We see that the only source of scatter is due to the best-fit parameter set $\boldsymbol{p}_0$, which varies with the noise realisation of the data. If we further assume that the curvature of the likelihood does not vary strongly as the best-fit position moves, only the last term in Eq. (\ref{eq:scatter_evidence}) is relevant for the statistical uncertainty in $\ln {\cal Z}$, while $D_{\rm KL}$ is robust to the scatter. Since $\ln {\cal Z} + D_{\rm KL} \approx \mbox{const.} - \chi^2(\boldsymbol{p}_0)/2$, $\ln S$ has identical noise properties to $\ln {\cal Z}$ under these assumptions.

Equipped with these considerations, we propose the following algorithm: (i) perform a single full likelihood analysis and determine fiducial evidence values, ${\cal Z}_{\rm fid}$; (ii) generate samples of the data from the likelihood; (iii) determine the maximum of the likelihood, or equivalently $\chi^2_{\rm min}$, for each sample\footnote{In practice, we obtain the minimum $\chi^2$ within the wide prior ranges of the parameters.}; and (iv) derive samples of the evidence via 
\eq{
\label{eq:zapprox}
\ln {\cal Z}_{{\rm approx},i} := \ln {\cal Z}_{\rm fid} - \frac{1}{2} \br{ \chi^2_{{\rm min},i} - \chi^2_{\rm min, fid} }\;.
}
Following this procedure with 100 samples results in the blue points shown in Fig.$\,$\ref{fig:zsamples} and the blue distribution in Fig.$\,$\ref{fig:rdistribution}. Apart from sampling noise in the tail, we recover the true distribution well, with the mean and variance in agreement within $\sim 10\,\%$. The change from a full exploration of the posterior, which typically runs in hours to days, to a maximisation of the likelihood, which usually takes minutes to hours, makes exploring the noise properties of the Bayesian evidence and its derived quantities feasible.

\section{Evidence samples from real data}
\label{sec:evidence_real}

When analysing real data, the true likelihood $\pr(\boldsymbol{d}'|M_{\rm true})$ found in Eq.~(\ref{eq:bayes_distribution}) needed to generate new copies of the data vector is unavailable.  Our best guess for this truth is the best-fitting model, which itself carries uncertainty as it is inferred from the data.  In this case we can make use of the posterior predictive distribution \citep[PPD;][]{gelman96}, $\pr(\boldsymbol{d}'|\boldsymbol{d},M_k)$, which yields new samples of the data $\boldsymbol{d}'$ for a given observation $\boldsymbol{d}$ assuming model $M_k$ (see \citealp{trotta07} for a very similar application of the PPD).
Averaging over all models using the posterior model probabilities $\pr(M_k|\boldsymbol{d})$ from Eq.~(\ref{eq:defbayesfactor}) then yields
\eqa{\pr(\boldsymbol{d}'|M_{\rm true}) \approx \pr(\boldsymbol{d}'|\boldsymbol{d}) = \sum_k \pr(\boldsymbol{d}'|\boldsymbol{d},M_k) \pr(M_k|\boldsymbol{d})\;.
}
The algorithm presented in Sect.$\,$\ref{sec:fastapprox} therefore only needs to be adjusted in Step (ii), where instead of generating data realisations from the true likelihood in the mock scenario, they are now produced from the PPD via random selection of a subset of posterior samples and evaluation of the likelihood at the parameter values corresponding to these samples.

In practice, we simplified the approach by choosing the model that yields the higher evidence to produce the PPD samples rather than full model averaging. If both models have similar evidence, this choice should have little impact; if the evidence ratio is large, the model with higher evidence is more accurate and/or more predictive (cf. the solid and hatched regions in Fig.~\ref{fig:sketch}).

\section{Application to KiDS-450 internal consistency}
\label{sec:application}

\begin{figure}
        \centering
        \includegraphics[width=\columnwidth]{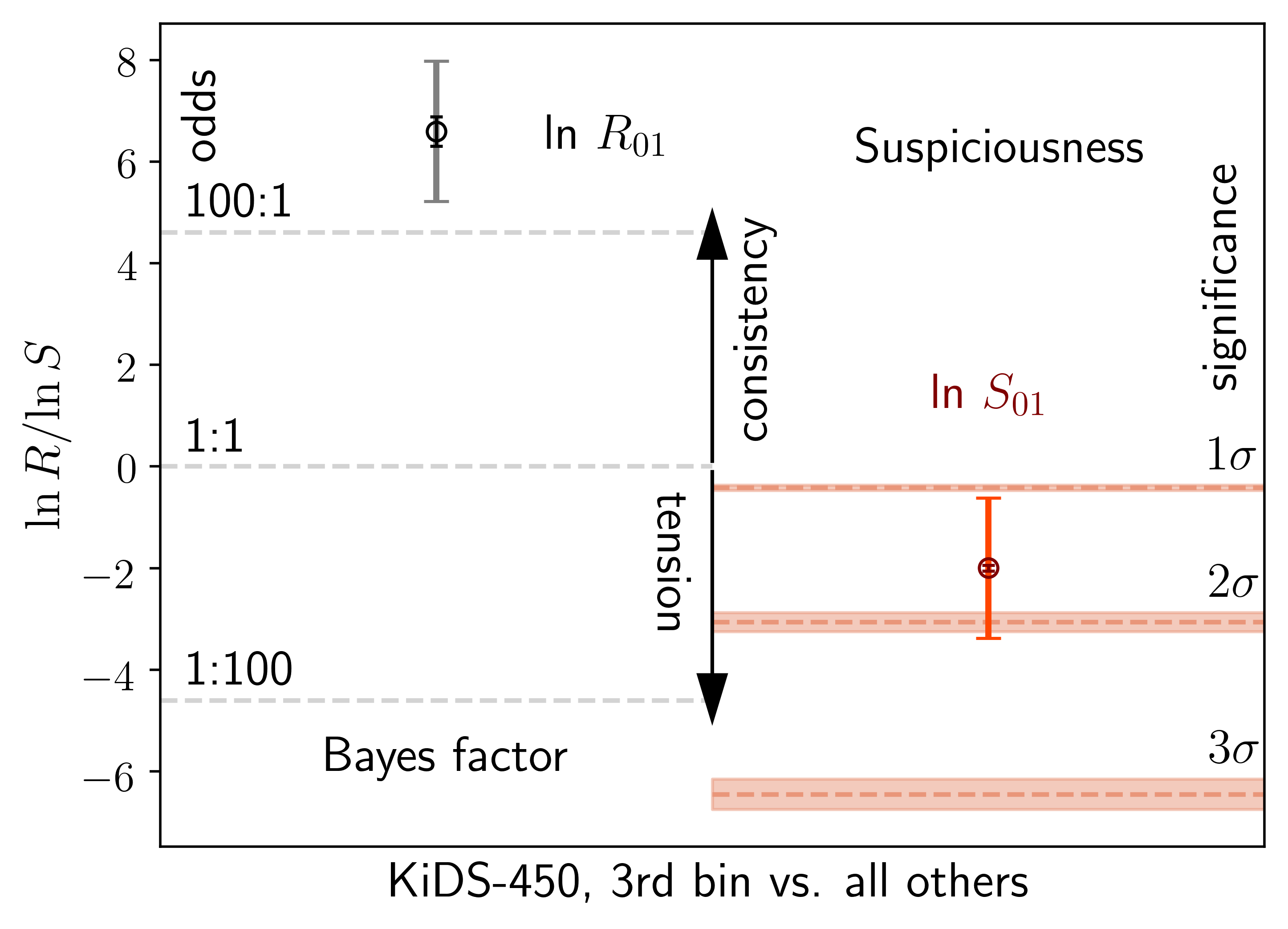}
    \caption{Tension statistics for the case of KiDS-450 internal consistency with respect to tomographic bin no. 3. Shown are the Bayes factor $\ln R_{01}$, with odds ratios, as well as the suspiciousness $\ln S_{01}$, with tension significance, in multiples of the width of an equivalent Gaussian, $\sigma$. The smaller red and black error bars are the errors associated with the nested sampling, while the larger orange and grey error bars are the statistical errors derived in this work. Red bands show the statistical uncertainty in determining the $\sigma$-levels for $\ln S_{01}$.}
        \label{fig:realdata}
\end{figure}

We then inserted the real KiDS-450 data vector into our analysis and generated ten PPD samples from the joint Model 0 as this yields significantly higher evidence than the split model. The derived standard deviations of $\ln R$ and $\ln S$ are shown in Fig.$\,$\ref{fig:realdata}. These statistical errors far exceed the typically quoted \lq method\rq\ errors, which derive from the finite sampling of the posterior. The interpretation of the suspiciousness acquires an additional, albeit smaller, source of error through the effective model dimension, $m_{\rm eff}$, that determines the $\sigma$-levels.

The noise in the tension statistics leads to a more conservative evaluation of discrepancies in the data. While the point estimate suggests  tension at $1.6\sigma$, this reduces to $1.1\sigma$ if we require that all but $16\,\%$ (i.e. the one-sided tail beyond $1\sigma$ of a normal distribution) of possible realisations of the data are discrepant by at least that level. Visually, this corresponds to the upper $1\sigma$ error of $\ln S$ almost touching the lower limit of the $1\sigma$ band in Fig.$\,$\ref{fig:realdata}.

\section{Benefits of data compression}

\textit{Planck} CMB data are at the centre of both current major tension controversies in cosmology. A practical obstacle to applying our formalism is the complexity of the \textit{Planck} temperature likelihood, which is assumed to be Gaussian only for $\ell>30$ and builds on pixelised sky maps on larger scales \citep{planck18_likelihood}. This makes drawing PPD samples challenging. However, \citet{prince19} recently showed that the low-$\ell$ likelihood can be efficiently compressed into two Gaussian-distributed band powers. They proceeded to apply maximal, linear compression (using the Multiple Optimised Parameter Estimation and Data or MOPED scheme, \citealp{tegmark97,heavens00}) to the full temperature likelihood and demonstrated it to be nearly lossless. This is not unexpected since the cosmological sampling parameters in CMB analyses are chosen to be close to linear and to  be Gaussian-distributed \citep{kosowsky02}.

There is an additional motivation to apply data compression: It can suppress scatter in the Bayesian evidence. Under the assumptions outlined in Sect.$\,$\ref{sec:fastapprox}, the statistical properties of $\ln {\cal Z}$ are driven by the distribution of $\chi^2(\boldsymbol{p}_0)$ (i.e. the minimum $\chi^2$; cf. Eq.$\,$\ref{eq:scatter_evidence}). If the data are approximately Gaussian and well fitted by a model whose parameters are close to linear, $\chi^2(\boldsymbol{p}_0)$ follows a $\chi^2$-distribution with $N_{\rm dof}=n-m$ degrees of freedom, so that ${\rm Var}(\ln {\cal Z}) = 2 N_{\rm dof}$. Data compression decreases $n$ and can yield $N_{\rm dof} \approx 0$ in the maximal case, that is, evidence becomes essentially noise-free because a good model with $n$ linear parameters perfectly fits $n$ compressed data. Appendix~\ref{app:gaussian_analytic} demonstrates this explicitly for the Gaussian case. 

This may seem paradoxical because compression can at best preserve information, raising the question of how it can facilitate a more precise determination of evidence.
In the context of Fig.~\ref{fig:sketch}, compression reduces the scatter between the model parameter and the data, so that for a given parameter the data vary little and thus the evidence is known precisely. Conversely, a broad likelihood and/or a high-dimensional data vector lead(s) to large variations in possible realisations of data. While this has no bearing on the posterior, and therefore on the information content, it increases the probability that a certain level of tension or model preference is owed to a particularly (un)lucky noise realisation of the data vector and does not reflect a physical trend. 

As a proof of concept, we adopted the \citet{prince19} approach, using the provided software\footnote{\texttt{https://github.com/heatherprince/planck-lite-py}}, and compressed the \textit{Planck} temperature anisotropy power spectra into the six cosmological parameters of a spatially flat $\Lambda$CDM model (nuisance parameters are marginalised over pre-compression). We then determined the $\chi^2_{\rm min}$ for the compressed real data, as well as for new data realisations generated from the compressed likelihood. We find an extremely small $\chi^2_{\rm min} $ ($\approx 1.4 \times 10^{-8}$) for the real data and similar values for the noise realisations, with a standard deviation of $4.4 \times 10^{-9}$. Hence, practically noise-free evidence measurements from \textit{Planck} are indeed possible.

\section{Conclusions}

We studied the impact on model comparison statistics if these are to be interpreted based on the ensemble of possible observations rather than a single observed realisation of the data.
In this setting they become noisy quantities, which affects binary decisions on signal detection, model selection, and tension between experiments.
Confirming earlier analytic arguments, we found standard deviations of order unity for the logarithm of the Bayes factor and the suspiciousness statistic, with substantially broader distributions in the case of strong discrepancies between the models under comparison. We expect these conclusions to apply to most, possibly all, informative tension metrics available in the literature as they typically depend on the maximum likelihood or $\chi^2$-like expressions.

We proposed a method to approximate the probability distribution of the evidence via repeated draws of mock data from the likelihood, with the maximum likelihood for each mock dataset then obtained, that will add negligible computation time to a full exploration of the posterior distribution. Conclusions drawn from noisy model comparison measures inevitably become more conservative, for example, the tension significance according to the suspiciousness for an internal consistency analysis of KiDS weak lensing data reduces from $1.6\sigma$ in the traditional approach to $1.1\sigma$ when scatter is accounted for. While in this application the two models under comparison were nested, our formalism and conclusions also hold for the more general case in which parameter spaces differ.

Finally, we demonstrated that data compression suppresses the impact of noisy data on the evidence, in the case of \textit{Planck} CMB constraints to negligible levels. In light of this, the following pre-processing steps are beneficial before any form of model comparison: (i) compress the data vector as much as possible as long as the compression is essentially lossless; and (ii) choose a parametrisation such that the model is close to linear in the parameters \citep[see e.g.][]{schuhmann16}, which increases the chances of achieving a near-perfect fit for any noise realisation of the data.

\begin{acknowledgements}
We thank C. Jenkins, J. Peacock, and R. Trotta for insightful discussions. We are also grateful to J. Peacock for a thorough review of the manuscript and to C. Heymans for alerting us to the work of Jenkins and Peacock.
BJ acknowledges the kind hospitality of IPMU where part of this work was carried out.
FK acknowledges support from the World Premier International Research Center Initiative (WPI), MEXT, Japan.
PL acknowledges STFC Consolidated Grant ST/R000476/1.\\

The KiDS analysis is based on observations made with ESO Telescopes at the La Silla Paranal Observatory under programme IDs177.A-3016,  177.A-3017,  177.A-3018,  179.A-2004,  298.A-5015, and on data products produced by the KiDS Consortium.

\end{acknowledgements}

\bibliographystyle{aa}
\bibliography{bibliography}

\begin{appendix} 

\section{The complete Gaussian case}
\label{app:gaussian_analytic}

If we assume that an experiment produces a single observation of $n$ data points\footnote{Equivalently, one could consider a data vector that has been averaged over multiple observations.} drawn from a Gaussian distribution about some true mean $\bar{\vec{d}}$ with covariance $\tens{C}$,
\begin{equation}
    \ln \pr(\vec{d}) = -\frac{1}{2}\ln|2\pi \, \tens{C}| - \frac{1}{2}(\vec{d}-\bar{\vec{d}})^\tau \tens{C}^{-1} (\vec{d}-\bar{\vec{d}})\,.
    \label{eqn:data_likelihood}
\end{equation}
In general we do not know $\bar{\vec{d}}$ but design a model, $M$, that parameterises the data by some function ${\vec f}(\vec{p})$, with $m\ll n$ parameters, in the hope that the true data are well approximated by our model. Assuming a given model, the likelihood becomes
\begin{equation}
    \ln \pr(\vec{d}|\vec{p},M) = -\frac{1}{2}\ln|2\pi \, \tens{C}| - \frac{1}{2} \bb{ \vec{d}-{\vec f}(\vec{p}) }^\tau \tens{C}^{-1} \bb{ \vec{d}-{\vec f}(\vec{p}) }.
    \label{eqn:likelihood}
\end{equation}
A likelihood can often be approximated as a Gaussian in the parameter space,
\begin{equation}
    \ln \pr(\vec{d}|\vec{p},M) = \ln L_\mathrm{max} - \frac{1}{2}(\vec{p}-\vec{\mu})^\tau \, \tens{\Sigma}^{-1}(\vec{p}-\vec{\mu}),
    \label{eqn:likelihood_approx}
\end{equation}
with mean $\vec{\mu}$ and covariance $\tens{\Sigma}$, and the corresponding log-evidence reads
\begin{equation}
    \ln {\cal Z} \equiv \ln \pr(\vec{d}|M) = \ln L_\mathrm{max}  + \ln \frac{\sqrt{|2\pi \, \tens{\Sigma}|}}{V_{\rm prior}}\,,
    \label{eqn:evidence_approx}
\end{equation}
where $V_{\rm prior}$ is the volume of a uniform prior fully encompassing the posterior.
One can make the link between Eqs.~(\ref{eqn:likelihood}) and (\ref{eqn:likelihood_approx}) explicit by assuming that we can model our function ${\vec f}$ as linear in the region of parameter space around $\vec{p}_*$ where the likelihood is significantly non-zero,
\begin{equation}
    {\vec f}(\vec{p}) \approx {\vec f}(\vec{p}_*) + \nabla {\vec f}(\vec{p}_*)\, (\vec{p}-\vec{p}_*) =: \hat{\vec{d}} + \tens{J}\, (\vec{p}-\vec{p}_*)\,,
    \label{eqn:function_approx}
\end{equation}
from which one can identify
\eqa{
    \ln L_\mathrm{max} &= - \frac{1}{2}\ln|2\pi \, \tens{C}| -\frac{1}{2}(\vec{d}-\hat{\vec{d}})^\tau \tilde{\tens{C}}^{-1} (\vec{d}-\hat{\vec{d}})  ~~~\mbox{and}  
    \label{eqn:evidence}\\
    \tens{\Sigma}^{-1} &= \tens{J}^\tau \tens{C}^{-1} \tens{J}; \qquad \vec{\mu} = \vec{p}_* + \tens{\Sigma} \tens{J}^\tau \tens{C}^{-1} (\vec{d}-\hat{\vec{d}})\,,
    \label{eqn:post_mean_var}
}
where we defined
\eqa{
    \tilde{\tens{C}}^{-1} &:=  \tens{C}^{-1} - \tens{C}^{-1} \tens{J} \tens{\Sigma} \tens{J}^\tau \tens{C}^{-1}.
}
As an aside, Eq.~(\ref{eqn:post_mean_var}) shows that noisy data realisations affect the posterior mean but not its covariance. In other words, while the posterior shape is unaffected, the distribution moves as a whole in parameter space with different realisations of the data.

From the above expressions we can immediately see that the evidence is quite a noisy statistic, driven by the second term in Eq.~(\ref{eqn:evidence}). Taking the variance of Eq.~(\ref{eqn:evidence_approx}) after inserting Eq.~(\ref{eqn:evidence}) and assuming that $\vec{d}$ follows the distribution of Eq.~(\ref{eqn:data_likelihood}) yields
\begin{equation}
\label{eqn:var_evidence}
    \mathrm{Var}(\ln {\cal Z}) = \frac{1}{2}\Tr \bb{ (\tilde{\tens{C}}^{-1} \tens{C})^2 } + (\bar{\vec{d}}-\hat{\vec{d}})^\tau \tilde{\tens{C}}^{-1} \tens{C} \tilde{\tens{C}}^{-1} (\bar{\vec{d}}-\hat{\vec{d}})\,.
\end{equation}
The first term here is equal to $\frac{1}{2}(n-m)$, and hence the variance in the raw evidence is large for $n \gg m$, even in the event of a good fit to the data (i.e. $\hat{\vec{d}} \approx \bar{\vec{d}}$). We also see that in the case of heavily compressed data, $n \sim m$, the evidence scatter reduces considerably.

To derive the expression~(\ref{eqn:var_evidence}), as well as some of the following equations, it is helpful to note that for a Gaussian-distributed variable $\vec{x}$ with covariance $\tens{C}$ centred on zero \citep{matrix_cookbook},
\eqa{
\nn
\langle (\vec{x}-\vec{a})^\tau \tens{A} (\vec{x}-\vec{a}) \rangle &=  \Tr \bb{\tens{A} \tens{C} } + \vec{a}^\tau \tens{A} ~~\vec{$ and$} \,   \\ \nn 
\mathrm{Cov} \left[ (\vec{x}-\vec{a})^\tau \tens{A} (\vec{x}-\vec{a}), \right.&\left. (\vec{x}-\vec{b})^\tau \tens{B} (\vec{x}-\vec{b}) \right] \\ 
&=  2\Tr \bb{ \tens{A}\tens{C} \tens{B}\tens{C} } + 4 \vec{b}^\tau \tens{B} \tens{C} \tens{A} \vec{a}\,,
}
where $\tens{A}$ and $\tens{B}$ are symmetric matrices, and $\vec{a}$ and $\vec{b}$ are arbitrary, non-stochastic vectors.

We are of course really interested in how model comparison (i.e. a difference in evidence) scatters with noisy data, so we introduced two models, one with with $\hat{\vec{d}}_1$ and $\tens{J}_1$ and one with $\hat{\vec{d}}_2$ and $\tens{J}_2$\footnote{In general, the models under comparison do not need to share any part of their parameter space, in which case the pivot $\vec{p}_*$ in Eq.~(\ref{eqn:function_approx}) could also differ. }, and asked what the variance in their evidence difference is.  Under the true distribution of Eq.~(\ref{eqn:data_likelihood}), we find that the log Bayes factor (under the same assumptions as in Eq.~\ref{eq:defbayesfactor}), $\ln R_{12} =\ln {\cal Z}_1 - \ln {\cal Z}_2$, has a mean
\begin{align}
    \langle \ln R_{12} \rangle =&
    \frac{1}{2}(\bar{\vec{d}}-\hat{\vec{d}}_2)^\tau \tilde{\tens{C}}_2^{-1} (\bar{\vec{d}}-\hat{\vec{d}}_2)
    -\frac{1}{2}(\bar{\vec{d}}-\hat{\vec{d}}_1)^\tau \tilde{\tens{C}}_1^{-1} (\bar{\vec{d}}-\hat{\vec{d}}_1)
    \nonumber\\
    &+\frac{1}{2}\Tr \bb{ \Delta } 
    + \ln \frac{\sqrt{|2\pi \, \tens{\Sigma}_1|} V_{\rm prior,2}}{\sqrt{|2\pi \, \tens{\Sigma}_2|} V_{\rm prior,1}}\,,
    \label{eqn:mean_ev_diff} 
\end{align}
(see also \citealp{lazarides04,heavens07} for similar, less general expressions) and a variance
\eqa{
    \label{eqn:var_ev_diff}
    \mathrm{Var}(\ln R_{12}) &= 
    \frac{1}{2}\Tr \bb{ \Delta^2 }
    + (\Delta\, \bar{\vec{d}} - \delta)^\tau \tens{C}^{-1}(\Delta\, \bar{\vec{d}} - \delta)\,,
}
where we defined
\eqa{
    \Delta &:= \tens{C}(\tilde{\tens{C}}_2^{-1} - \tilde{\tens{C}}_1^{-1})\,; \qquad \delta := \tens{C}(\tilde{\tens{C}}_2^{-1}\hat{\vec{d}}_2- \tilde{\tens{C}}_1^{-1}\hat{\vec{d}}_1)\,.
}
The mean in Eq.~(\ref{eqn:mean_ev_diff}) has three portions: a set of misfit terms on the first line, a constant trace term equal to $\frac{1}{2}(m_1-m_2)$, and an Occam factor. The trace contribution can be understood as a typically small modification of the Occam factor.

In the variance (Eq.~\ref{eqn:var_ev_diff}), there is a trace term that is roughly the dimensionality of the parameter space(s) $\le\frac{1}{2}(m_1+m_2)$, as well as a data misfit term. The trace term is always present, representing the \lq order unity\rq\ term for the general Gaussian case, but can reduce towards zero (via a cross-term that is dependent on both models), getting closer to zero the more similar the two model parametrisations (as quantified by $\tens{J}$) are to each other. As opposed to the variance of the evidence (cf. Eq.~\ref{eqn:var_evidence}), the trace term in the variance of the Bayes factor does not depend on $n$, so if $n \gg m$, the scatter in $R_{12}$ is significantly smaller than the scatter in either ${\cal Z}_1$ or ${\cal Z}_2$ if the data is well fitted. This is the situation we encountered in Fig.~\ref{fig:zsamples}.

The second term can be small if the models are good, but can also become arbitrarily large, which corresponds to the scatter seen in Fig.~\ref{fig:rdistribution}. It should be noted that in the event of large misfits, the mean and variance are both of the same order, which gives a Poisson-type evidence error associated with measurement noise. This is reassuring as it means the evidence in theory becomes relatively less noisy the larger it becomes. We note that if Eq.~(\ref{eqn:function_approx}) is a reasonable approximation, provided one can compute (by numerical derivatives or otherwise) the Jacobian $\tens{J}$, one may use Eq.~(\ref{eqn:var_ev_diff}) to evaluate the expected scatter, for example by employing the observed data vector as an estimate of the true $\bar{\vec{d}}$.

Finally, we offer some illustration for the use of information regarding the sampling distribution of the Bayes factor (such as the variance of Eq.~\ref{eqn:var_ev_diff}) by invoking the popular analogy with betting odds. Rather than placing one's bets based on a single measure of the odds conditioned on some observation, it is beneficial to take the scatter of these odds into account, even if the scatter is built upon an imperfect model and noisy data. The scatter might indicate that a different outcome has a substantial probability, and hence it would be wise to invest one's money more cautiously.

\section{Details of the likelihood analysis}
\label{app:kids}

For most of our analyses we employed simulated and real data from the KiDS-450 analysis \citep{Hildebrandt2017}\footnote{The data are publicly available at \url{http://kids.strw.leidenuniv.nl/sciencedata.php}.}. We also adopted their five-parameter $\Lambda$CDM cosmological model with spatially flat geometry and used the same set of priors. The sample parameters are the amplitude of the primordial power spectrum $\ln (10^{10} A_{\rm s})$, the current value $h$ of the Hubble parameter divided by $100 \, {\rm km\,s^{-1}\,Mpc^{-1}}$, the cold dark matter density $\Omega_{\rm cdm}h^2$, the baryonic matter density $\Omega_{\rm b}h^2$, and the power-law exponent of the primordial power spectrum $n_{\rm s}$. In addition to these key cosmological parameters, we varied the free amplitude parameters of the intrinsic alignment and baryon feedback models,$A_{\rm IA}$ and $A_{\rm bary}$. The implementation of the inference pipeline is that presented in \citet{Koehlinger2019}\footnote{Likelihood pipelines available in {\scriptsize MONTE PYTHON}, \url{https://github.com/brinckmann/montepython_public} \citep{Audren2013, Brinckmann2018}, and from \url{https://github.com/fkoehlin/montepython_2cosmos_public}.}, which is independent of, but in excellent agreement with, the analysis of \citet{Hildebrandt2017}.

We opted for nested sampling \citep{Skilling2006} to explore the posterior distribution as the most efficient way to simultaneously evaluate high-dimensional likelihoods and calculate Bayesian evidence. To avoid significant algorithm-induced scatter in the evidence values, we checked three variants of nested sampling algorithms for their consistency. We use {\scriptsize MULTINEST}\footnote{Version 3.8 from \url{http://ccpforge.cse.rl.ac.uk/gf/project/multinest/}} \citep{Feroz2008, Feroz2009, Feroz2013}  and
{\scriptsize POLYCHORD}\footnote{Version 1.16 from \url{https://github.com/polychord/polychordlite}} \citep{Handley2015a, Handley2015b}, which primarily differ in the key step of how new \lq live\rq\ sampling points are drawn at each likelihood contour. Moreover, we considered an importance-sampled determination of the evidence in {\scriptsize MULTINEST} that utilises the full set of generated sample points and can achieve higher accuracy \citep{Feroz2013}.
For {\scriptsize MULTINEST,} 1000 live points were used with a sampling efficiency of 0.3 and a final error tolerance on the log-evidence of 0.1. Live points for {\scriptsize POLYCHORD} runs were 25 times the number of parameters (seven for Model 0; 14 for Model 1) with a final error tolerance on the log-evidence of 0.001.

\begin{figure}
        \centering
        \includegraphics[width=\columnwidth]{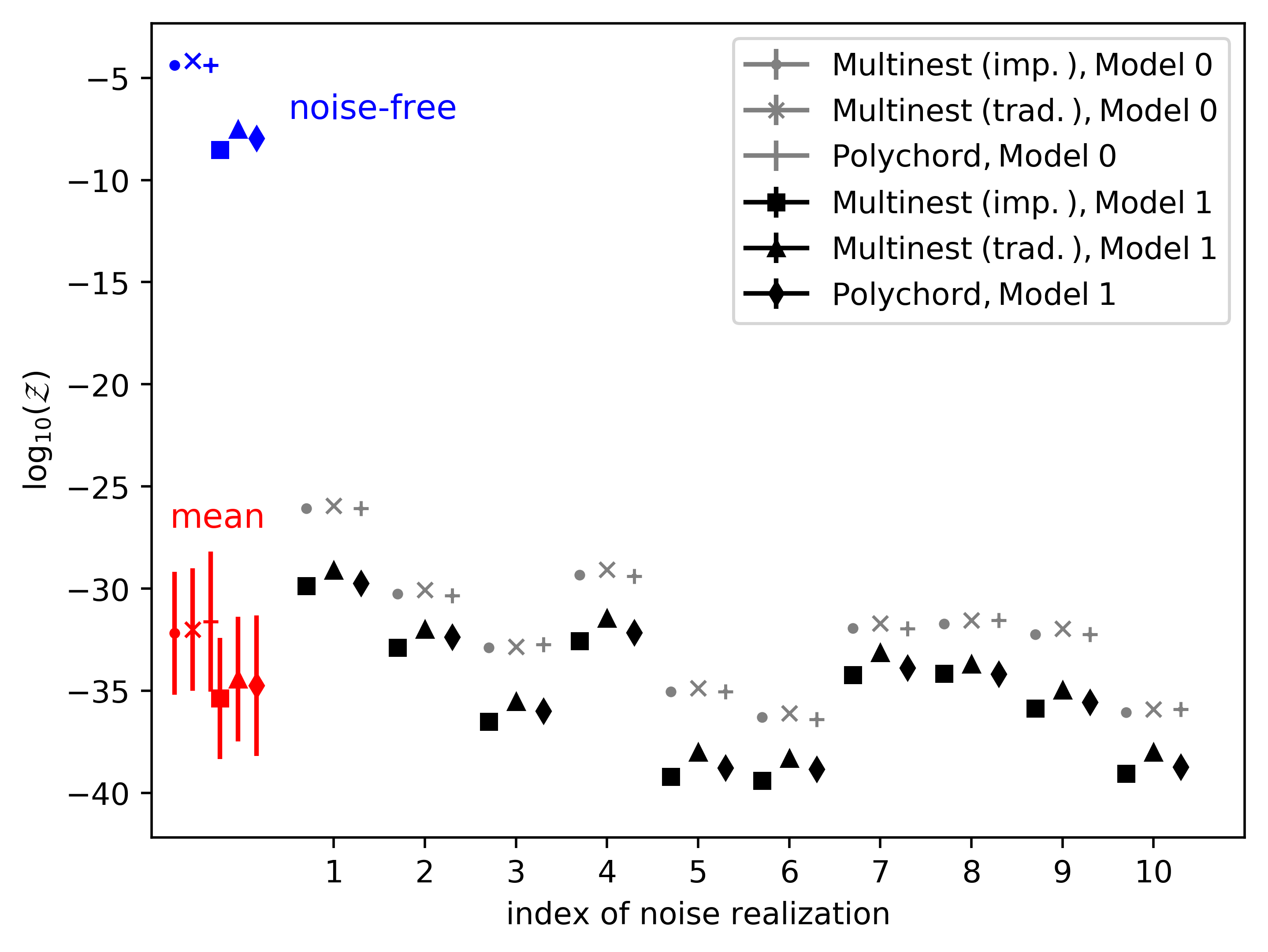}
    \caption{Comparison of sampler outputs. Black points correspond to the evidence of the joint analysis (Model 0), while grey points correspond to split analysis (Model 1) for ten noise realisations measured from the traditional or importance-sampled approaches of {\scriptsize MULTINEST}, as well as from {\scriptsize POLYCHORD}. Red points display the mean and standard deviation over these realisations. Blue points show results for a noise-free data vector.}
        \label{fig:comp_evid_PC_MN}
\end{figure}

Figure$\,$\ref{fig:comp_evid_PC_MN} shows evidence values for a KiDS-like noise-free simulated data vector, as well as for ten realisations with noise included. It is evident that in all cases, and for both the joint and split cosmological models, the three nested sampling variants agree very well with one another, with the residual scatter at a small fraction of the statistical errors. Our {\scriptsize MULTINEST} and {\scriptsize POLYCHORD} settings were optimised to yield accurate evidence. However, we note that evidence values are faithfully recovered as soon as the bulk of the posterior is explored, while credible regions of the parameters as well as the effective dimension (see Eq.$\,$\ref{sec:inf_model_comp}) are sensitive to the tails of the distribution. Therefore, when these tail-sensitive quantities are required in high-stakes real-data applications, we recommend increasing the accuracy settings of the nested sampling runs.

The $\chi^2$ minimisation for the approximate method was performed with the built-in {\scriptsize MONTE PYTHON} maximum likelihood determination, with a precision tolerance of $10^{-9}$ on the log-likelihood. With this setup, a minimisation run consumes about 500 times less wall-clock time than full, parallelised sampling on high-performance computing infrastructure.

\end{appendix}

\end{document}